\documentclass[twocolumn]{aastex6}

\RequirePackage{lineno}
\usepackage{amsmath}
\usepackage{amssymb}
\usepackage{hyperref}
\usepackage{amsthm}
\usepackage{natbib,aasdefs,url,bm}
\usepackage{array}
\usepackage{float}
\usepackage{graphicx}
\usepackage{subfigure}
\usepackage{color}
\usepackage{lineno}

\newcounter{ichi}
\newcounter{ni}
\newcounter{san}
\newcounter{yon}
\def\be{\begin{linenomath*}\begin{equation}}
\def\ee{\end{equation}\end{linenomath*}}
\def\ba{\begin{eqnarray}}
\def\ea{\end{eqnarray}}

\makeatletter
\newcommand\mytuple[1]{%
  \@tempcnta=0
  \bigl\langle
  \@for\@ii:=#1\do{%
    \@insertbreakingcomma
    \textit{\@ii}
  }%
  \bigr\rangle
}
\def\@insertbreakingcomma{%
  \ifnum \@tempcnta = 0 \else\,,\ \linebreak[1] \fi
  \advance\@tempcnta\@ne
}
\makeatother

\preprint{}
\shorttitle{G$\rm\MakeLowercase{e}$V Signatures of Short GRBs in AGNs}

\begin{document}
\title{$\rm G\MakeLowercase{e}V$ SIGNATUREs OF SHORT GAMMA-RAY BURSTS IN ACTIVE GALACTIC NUCLEI}
\author{Chengchao Yuan\altaffilmark{1}}\email{cxy52@psu.edu}
\author{Kohta Murase\altaffilmark{1,2}}\email{murase@psu.edu}
\author{Dafne Guetta\altaffilmark{3}}
\author{Asaf Pe'er\altaffilmark{4}}
\author{Imre Bartos\altaffilmark{5}}
\author{P\'eter M\'esz\'aros\altaffilmark{1}}
\altaffiltext{1}{Department of Physics, Department of Astronomy \& Astrophysics, Center for Multimessenger Astrophysics, Institute for Gravitation and the Cosmos, The Pennsylvania State University, University Park, PA 16802, USA}
\altaffiltext{2}{Center for Gravitational Physics, Yukawa Institute for Theoretical Physics, Kyoto University, Kyoto, Kyoto 606-8502, Japan}
\altaffiltext{3}{Department of Physics, Ariel University, Ariel, Israel}
\altaffiltext{4}{Department of Physics, Bar-Ilan University, Ramat-Gan 52900, Israel}
\altaffiltext{5}{Department of Physics, University of Florida, PO Box 118440, Gainesville, FL 32611-8440, USA }
\date{\today}
\begin{abstract}
The joint detection of gravitational waves and the gamma-ray counterpart of a binary neutron star merger event, GW170817, unambiguously validates the connection between short gamma-ray bursts and compact binary object (CBO) mergers. We focus on a special scenario where short gamma-ray bursts produced by CBO mergers are embedded in disks of active galactic nuclei (AGN),
and we investigate the $\gamma$-ray emission produced in the internal dissipation region via synchrotron, synchrotron self-Compton and external inverse-Compton (EIC) processes. In this scenario, isotropic thermal photons from the AGN disks contribute to the EIC component. We show that a low-density cavity can be formed in the migration traps, leading to the embedded mergers producing successful GRB jets. We find that the EIC component would dominate the GeV emission for typical CBO mergers with an isotropic-equivalent luminosity of $L_{j,\rm iso}=10^{48.5}~\rm erg~s^{-1}$ which are located close to the central supermassive black hole.
Considering a long-lasting jet of duration $T_{\rm dur}\sim10^2-10^3$ s, we find that 
the future CTA will be able to detect its $25-100$ GeV emission out to a redshift $z=1.0$. {In the optimistic case, it is possible to detect the on-axis extended emission simultaneously with GWs within one decade using MAGIC, H.E.S.S., VERITAS, CTA, and LHAASO-WCDA. Early diagnosis of prompt emissions with \emph{Fermi}-GBM and HAWC can provide valuable directional information for the follow-up observations.}
\end{abstract}
\maketitle

\section{Introduction}
As one of the most luminous and energetic phenomena in the universe, gamma-ray bursts (GRBs) have fueled a vibrant field of astrophysics research for several decades. Based on the duration of the bursts, the population can be divided into two subclasses, long GRBs and short GRBs, which are thought to arise from different progenitors. The general view is that short GRBs result from compact binary object (CBO) mergers \citep[e.g.,][]{1986ApJ...308L..43P,1989Natur.340..126E,1992ApJ...397..570M,1992ApJ...395L..83N,lee2007progenitors,2014ARA&A..52...43B}, such as binary neutron star mergers and potentially NS-black hole mergers, whereas long GRBs are generated during the death of massive stars \citep[e.g.,][]{woosley1993gamma,paczynski1998gamma,popham1999hyperaccreting,macfadyen1999collapsars,meszaros2006gamma,hjorth_bloom_2012}. In 2017, the coincident detection of gravitational waves (GWs) and the corresponding electromagnetic counterpart from the binary neutron star merger GW170817, located in the host galaxy NGC 4933, marked a triumph of multimessenger astronomy \citep{2017PhRvL.119p1101A,2017Natur.551...85A,2017apjl551...85A,2017apj2l551...85A}. The spatial and temporal association between GW170817 and the gamma-ray burst GRB 170817A also consolidates the theory that CBO mergers are the origin of short GRBs. Extensive efforts have shown that the broadband emission is consistent with a relativistic jet viewed from an off-axis angle \citep{2017apj2l551...85A,2017ApJ...848L..14G,2017Sci...358.1579H,2017ApJ...848L..15S,2017Natur.551...71T,2018NatAs...2..751L,2018PhRvL.120x1103L,2018Natur.561..355M,2018Natur.554..207M,2019MNRAS.488.1416G,2019MNRAS.487.4884I}. 
Moreover, \cite{2019ApJ...887L..16K} investigated the upscattered cocoon emission as the source of the $\gamma$-ray counterpart. The long-lasting high-energy signatures of the central engine left after the coalescence was studied in \cite{2018ApJ...854...60M}. 

Alternatively, unlike in the case of GW170817, one can expect a sub-population of short GRBs which occur in the accretion disks of AGNs. Studies of the CBO formation and evolution in AGN disks demonstrate that hierarchical mergers of embedded binary black hole systems are promising for reconstructing the parameters of LIGO/VIRGO detected mergers \citep{2020ApJ...890L..20G,2020arXiv201009765S,2021AAS...23723402B,2021MNRAS.505.2170T}. These mergers can harden the black hole mass distribution \citep{2019ApJ...876..122Y,2019PhRvL.123r1101Y,2020ApJ...898...25T,2021ApJ...908..194T} as well. 
\cite{2020ApJ...901L..34Y} pointed out that mergers involving neutron stars, such as GW190814 and GW190425, could also arise in AGN disks. 
Recent progress on the optical counterpart to GW190521 could support this \citep{2020PhRvL.124y1102G}, although the confirmation needs further observations \citep{2021CQGra..38w5004A}. 
\cite{2021ApJ...906L...7P} systematically studied the electromagnetic signatures of both long GRBs and short GRBs in AGN disks and discussed the conditions for shock breakout. \cite{2021arXiv210706070Z} and \cite{2021ApJ...911L..19Z} focused more on the neutrino production of embedded explosions. However, \citet{2021ApJ...916..111K} showed that CBO environments are likely to be thin because of outflows that are common in super-Eddington accretion.   

In this work, we study $\gamma$-ray emission from short GRBs that are embedded in AGN disks. Inside the accretion disk, the embedded objects can migrate towards a migration trap due to angular momentum exchange via the torques originated from the disk density perturbations. {At the migration trap, the gas torque changes sign, and an equilibrium is achieved as the outwardly migrating objects meet inwardly migrating objects.} Numerical calculations show that compact binaries are typically formed near the migration trap at distances around $R_d\sim20-300R_{\rm S}$ to the central supermassive black hole \citep[SMBH,][]{2016ApJ...819L..17B}, where $R_{S}=2GM_{\star}/c^2$ is the Schwarzschild radius. Employing one-dimensional N-body simulations, \cite{2020ApJ...898...25T} obtained a more distant location for typical mergers at $\sim10^{-2}{-10^{-1}}$ pc (${\sim10^3-10^{4}R_{\rm S}}$ for a SMBH with mass $M_{\star}=10^8M_\odot$). We concentrate on the embedded GRBs with distances $R_d\sim10-10^3R_{\rm S}$. {We will show that AGN disks would not influence the $\gamma$-ray emission if the CBO mergers happen further outside in the disk. We also note that $R_d=10R_{\rm S}$ is an extreme case where the population is stringently limited.} The outflows from the binary systems with {super-Eddington accretion rates} are expected to form a low-density cavity-like structure before the merger occurs \citep{2021ApJ...916..111K}. Within such a cavity a successful GRB jet is likely to develop, since the ambient gas density is not sufficiently high to stall the jet, in contrast to the choked-jet case discussed in \cite{2021ApJ...911L..19Z}. 

In GRB theories, EIC processes can be important when seed photons in the external regions or late/early-time dissipation processes can be efficiently upscattered to the GeV-TeV bands by accelerated electrons \citep[e.g.,][]{2011ApJ...732...77M,2012ApJ...755...12V,2019ApJ...887L..16K}. {The EIC scenario can be used to explain the observed very-high-energy (VHE) emission from GRBs \citep[e.g.][]{2021ApJ...920...55Z,2021ApJ...908L..36Z}.} In the present case, the disk black body emission provides an appropriate supply of thermal photons to the short GRB jets. 

Adopting a thin-disk model, we derive the conditions for cavity formation and calculate disk photon spectra in \S\ref{sec:disk}. In \S\ref{sec:electrons}, we numerically solve the steady-state transport equation to obtain the electron distribution inside the jet. In \S\ref{sec:GAMMAspectra}, we calculate the synchrotron, synchrotron self-Compton (SSC) and EIC components. The effects of $\gamma\gamma$ absorption in the AGN disk and electromagnetic cascades are also taken into account. {We also present the detection perspectives for the \emph{Fermi} Large Area Telescope (\emph{Fermi}-LAT) and the VHE $\gamma$-ray facilities, such as the Major Atmospheric Gamma Imaging Cherenkov (MAGIC), the High Energy Stereoscopic System (H.E.S.S.), the Very Energetic Radiation Imaging Telescope Array System (VERITAS), the Cherenkov Telescope Array (CTA), and the water Cherenkov detector array in the Large High Altitude Air Shower Observatory (LHAASO-WCDA)}, in \S\ref{sec:detection}. {The prompt emissions are discussed in \S\ref{sec:prompt}.} We summarize and discuss the results in \S\ref{sec:summary}. 

Throughout the paper, we use the notation $Q_x=Q/10^x$, and physical quantities are written in CGS units unless otherwise specified. Quantities with the prime symbol, e.g., $Q'$, are written in the jet comoving frame. We use the symbol $F[a,b,c,...]$ to represent the value of a function $F$ evaluated at the point $(a,b,c,...)$.

\begin{figure*}\centering
\includegraphics[width=0.99\textwidth]{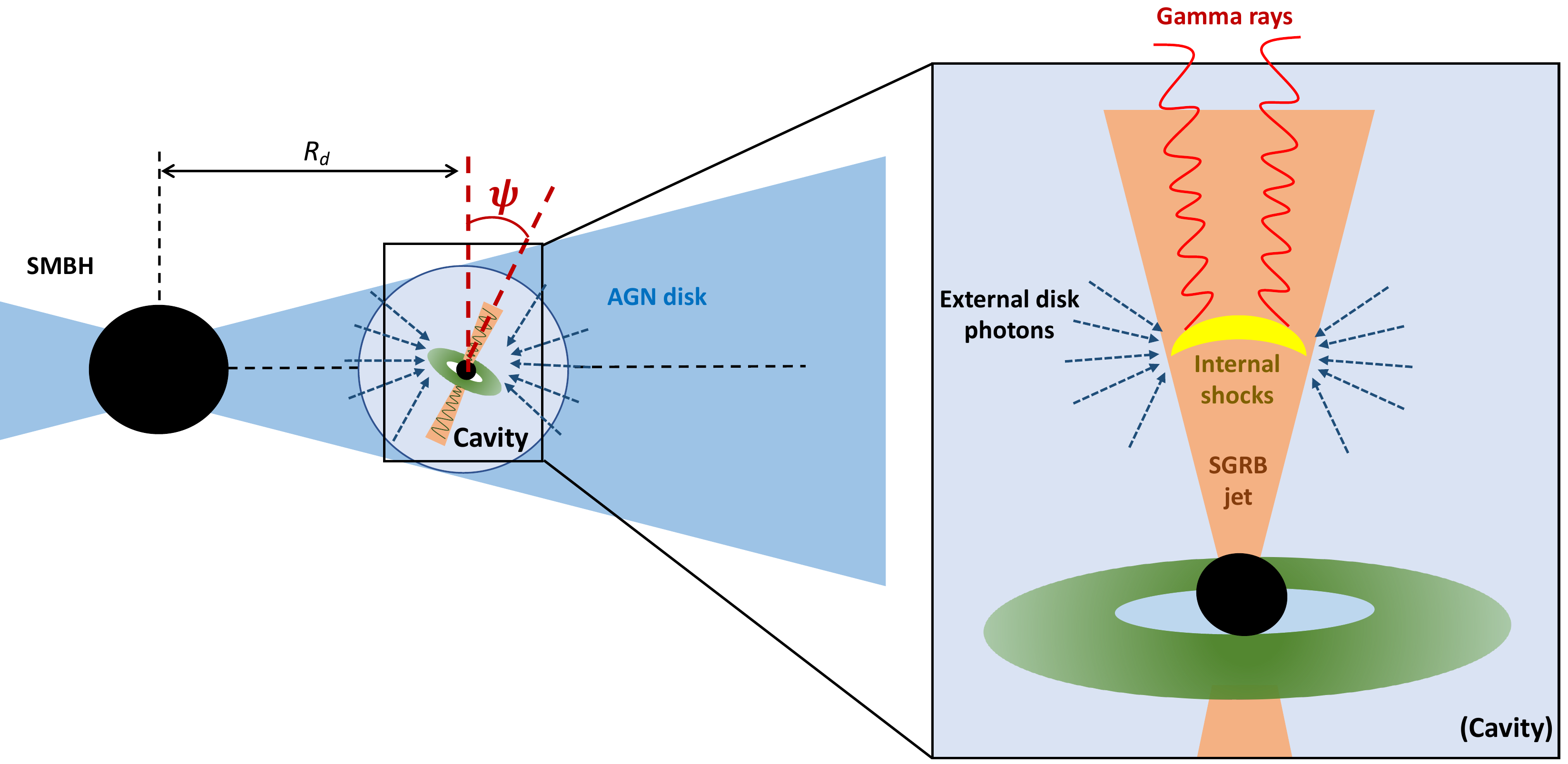}
\caption{Schematic {picture} of the CBO mergers embedded in AGN disks. A cavity is formed due to the powerful outflows from the circumbinary disk. In this configuration, $\psi$ represents the angle between the CBO orbital plane and the AGN disk, and $R_d$ is the distance between the CBO and the central SMBH. Non-thermal electrons accelerated in the internal dissipation region are responsible for the production of $\gamma$-rays. These electrons can upscatter the disk photons, leading to the EIC emission.}
\label{fig:schematic}
\end{figure*}


\section{Cavity formation and disk photon spectra}\label{sec:disk}
In this section we derive the conditions for the formation of a low-density cavity around the CBO, following the treatment in \cite{2021ApJ...916..111K}, and model the AGN disk temperature distribution assuming a steady thin disk. 

\subsection{Cavity formation}
For a thin AGN disk with an aspect ratio $h_{\rm AGN}=H_{\rm AGN}/R_d\sim0.01$ surrounding a SMBH with mass $M_{\star}=10^8M_{\star,8}M_\odot$, we write down the accretion rate onto the SMBH and the radial drift velocity $v_R$ as, respectively, $\dot M_{\star}=\dot m_{\star}L_{\rm Edd,\star}/c^2\simeq1.4\times10^{25}~\dot m_\star M_{\star,8}~{\rm g~s^{-1}}$ and $v_{R}=\nu/R_d\approx\alpha h_{\rm AGN}^2v_K\simeq2.1\times10^4~\alpha_{-1}h_{\rm AGN,-2}^2\mathcal R_{2}^{-1/2}~{\rm cm~s^{-1}}$ \citep{2002apa..book.....F}, where $H_{\rm AGN}$ is the scale height of the AGN disk, $\alpha\sim0.1$ is the viscous parameter, $\nu$ is the kinematic viscosity, $v_K=\sqrt{GM_\star/R_d}$ is the Kepler velocity, $R_d$ is the distance between the CBO and the central SMBH, the dimensionless parameter $\mathcal R$ is defined as $\mathcal R\equiv R_d/R_{\rm S}$, and $L_{\rm Edd,\star}$ stands for the Eddington luminosity. The surface density for a stable disk can then be written as $\Sigma_{\rm AGN}=\dot M_\star/(2\pi R_d v_R)\simeq3.6\times10^4~\dot m_\star M_{\star,8}\mathcal R_{2}^{-1/2}\alpha_{-1}^{-1}h_{\rm AGN,-2}^{-2}~{\rm g~cm^{-2}}$.
When a CBO is present in the AGN disk, the surface density is perturbed, and a density gap will appear bracketing the binary's orbit around the SMBH \citep{2015MNRAS.448..994K}. For a typical short GRB progenitor, we expect the total mass of the binary system to be $M_{\rm CBO}\lesssim10M_\odot$. In this case $\Sigma_{\rm CBO}\approx\Sigma_{\rm AGN}$ is a good approximation to the surface density of the AGN disk at the binary's position \citep{2021ApJ...916..111K}. We obtain the disk gas density in the vicinity of the CBO
\begin{linenomath*}\begin{equation}\begin{split}
\rho_{\rm CBO}=\frac{\Sigma_{\rm CBO}}{2H_{\rm AGN}}\simeq&6.1\times10^{-10}~\dot m_\star M_{\star,8}\\
&\times\mathcal R_{2}^{-3/2}\alpha_{-1}^{-1}h_{\rm AGN,-2}^{-3}~{\rm g~cm^{-3}}, 
\end{split}
\label{eq:CB_density}
\end{equation}\end{linenomath*}
{and the disk magnetic field 
\begin{linenomath*}\begin{equation}
\begin{split}
B_{d}&=\sqrt{8\pi\beta^{-1}(\rho_{\rm CBO}/m_p)k_BT_d}\\
&\simeq2.1\times10^2 ~\beta_{0.48}^{-1/2}\dot m_\star^{1/2} M_{\star,8}^{1/2}
 R_{2}^{-3/4}\alpha_{-1}^{-1/2}\\
 &~\times h_{\rm AGN,-2}^{-3/2}T_{d,5}^{1/2}~{\rm G},
    \end{split}
    \label{eq:B_d}
\end{equation}\end{linenomath*}
where $\beta\sim3-30$ is define as the ratio of the plasma pressure to the magnetic pressure and $T_{d}$ is the disk temperature.} Henceforth, the sub-index `CBO' will be used to stand for quantities describing CBOs. 

We estimate the accretion rate of the CBO to be $\dot M_{\rm CBO}\approx\eta_{\rm CBO}\dot M_{\star}\simeq1.4\times10^{24}~\dot m_\star M_{\star,8}\eta_{\rm CBO,-1}~{\rm erg~s^{-1}}$, where $\eta_{\rm CBO}$ is the ratio of the CBO accretion rate to the SMBH accretion rate. This approximation is justified in \cite{2021ApJ...916..111K}. We find that the accretion is highly super-Eddington, e.g., $\dot m_{\rm CBO}=\dot M_{\rm CBO}c^2/L_{\rm Edd,CBO}\simeq10^6\dot m_\star M_{\star,8}M_{\rm CBO,1}^{-1}\eta_{\rm CBO,-1}$, and expect a wind bubble to be produced by the strong radiation-driven outflows \citep[e.g.,][]{2009PASJ...61L...7O,jiang2014global,2014MNRAS.439..503S}. The bubble's expansion in a uniform medium can be described by the formula $r_B\approx0.88(L_wt^3/\rho_{\rm CBO})^{1/5}$ \citep[e.g.,][]{1977ApJ...218..377W,1992ApJ...388...93K}, where $r_B$ is the bubble radius, $L_w=\eta_w\dot M_{\rm CBO}v_w^2\simeq1.4\times10^{42}~\dot m_\star M_{\star,8}\eta_{\rm CBO,-1}\eta_wv_{w,9}^2~{\rm erg~s^{-1}}$ and $v_w\sim10^9v_{w,9}~\rm cm~s^{-1}$ is the outflow velocity. Since the accretion is highly super-Eddington, the factor $\eta_w$ can reach $\sim90-100\%$ \citep{2015ApJ...806...93J,2018PASJ...70..108K}. However, we use a conservative value $\eta_w\sim0.3~\eta_{w,-0.5}$ \citep{2014ApJ...796..106J}. Equating the bubble radius $r_B$ to $H_{\rm AGN}/\cos\psi$, we obtain the timescale to create a cavity reaching the approximate boundary of the AGN disk along the direction of the GRB jet,
\begin{linenomath*}\begin{equation}\begin{split}
    t_{\rm cav}&\approx1.2\left(\frac{\rho_{\rm CBO}H_{\rm AGN}^5}{L_w\cos^5\psi}\right)^{1/3}\\
    &\simeq4.0\times10^5~(\cos\psi)^{-5} \mathcal R_2^{7/6}\alpha_{-1}^{-1/3}h_{\rm AGN,-2}^{2/3}\\
    &~~~\times\eta_{\rm CBO,-1}^{-1/3}\eta_{w,-0.5}^{-1/3}v_{w,9}^{-2/3}~{\rm s},
    \end{split}
    \label{eq:cavity_timescale}
\end{equation}\end{linenomath*}
where $\psi$ is the angle between binary orbital plane and the AGN disk (see the schematic picture in Fig. \ref{fig:schematic}). {One possible caveat is that we have assumed a spherical outflow to derive the cavity timescale, equation \ref{eq:cavity_timescale}. \cite{2014MNRAS.439..503S} pointed out that the outflow is concentrated in a wide-angle funnel that surrounds the jet if the accretion rate is highly super-Eddington. In the following text, we will continue using the spherical cavity timescale for simplicity to obtain sufficient conditions for the cavity formation. This would give a reasonable approximation given that the outflow is non-relativistic.}

The formation of a cavity for a CBO located at $R_d$ before the merger occurs requires
\begin{linenomath*}\begin{equation}
    t_{\rm cav}\lesssim\min\left[t_{\rm gw},~t_{\rm mig},~t_{\rm vis}\right],
    \label{eq:cavity_condition}
\end{equation}\end{linenomath*}
where $t_{\rm gw}$, $t_{\rm mig}$, $t_{\rm vis}$ are binary merger, migration and AGN disk viscosity timescales, respectively. We write down the timescales for an equal-mass binary explicitly as
\begin{linenomath*}\begin{equation}
    \begin{split}
        t_{\rm gw}&=\frac{5}{128}\frac{\dot m_{\rm CBO}^4}{A_{\rm in}^4}\frac{GM_{\rm CBO}}{c^3}\\
        &\simeq1.9\times10^{14}~\dot m_{\rm CBO,6}^4A_{\rm in,1}^{-4}M_{\rm CBO,1}~{\rm s},\\
        t_{\rm mig} &=\frac{h_{\rm AGN}^2M_\star^2}{ M_{\rm CBO}R_dv_K\Sigma_{\rm AGN}}\\
        &\simeq 1.47\times10^{14}~\alpha_{-1}h_{\rm AGN,-2}^4M_{\star,8}^{1/2}M_{\rm CBO,1}^{-1}\dot m_\star^{-1}~\rm s,\\
        t_{\rm vis} &=\frac{R_d}{\alpha h_{\rm AGN}^2v_K}\\
        &\simeq1.39\times10^{11}~\alpha_{-1}^{-1}h_{\rm AGN,-2}^{-2} \mathcal  R_{2}^{3/2}M_{\star,8}^{-1/2}{~\rm s},
    \end{split}
    \label{eq:disk_timescales}
\end{equation}\end{linenomath*}
where $A_{\rm in}\sim10$ is the ratio of the inner edge of the circumbinary disk surrounding the CBO and the major axis of the binary's orbit \citep{2013MNRAS.434.1946N}. We define a critical angle $\psi_{c}$ above which the condition described by equation \ref{eq:cavity_condition} is no longer satisfied, and obtain
\begin{linenomath*}\begin{equation}
    \psi_c\simeq\frac{\pi}{2}-\max\left[h_{\rm AGN},~0.076\mathcal R_{2}^{-1/15}h_{\rm AGN,-2}^{8/15}M_{\star,8}^{1/10}\right].
    \label{eq:critical_angle}
\end{equation}\end{linenomath*}
In the equation above, the dependence on the parameters $\alpha,~\eta_w,~\eta_{\rm CBO}$ and $v_w$ are not shown, to simplify the notation. Varying $\mathcal R$ in the fiducial range $10-10^3$, we estimate the critical angle $\psi_c\simeq85.6^\circ$ and find that $\psi_c$ depends very weakly on $\mathcal R$ and $M_\star$. This result supports the argument that in most cases a cavity surrounding the CBO is unavoidable {and the jet is not choked}, except if the binary orbital plane is perpendicular to the AGN disk \citep{2021ApJ...916..111K}.
  
 \begin{figure*}\centering
\includegraphics[width=0.49\textwidth]{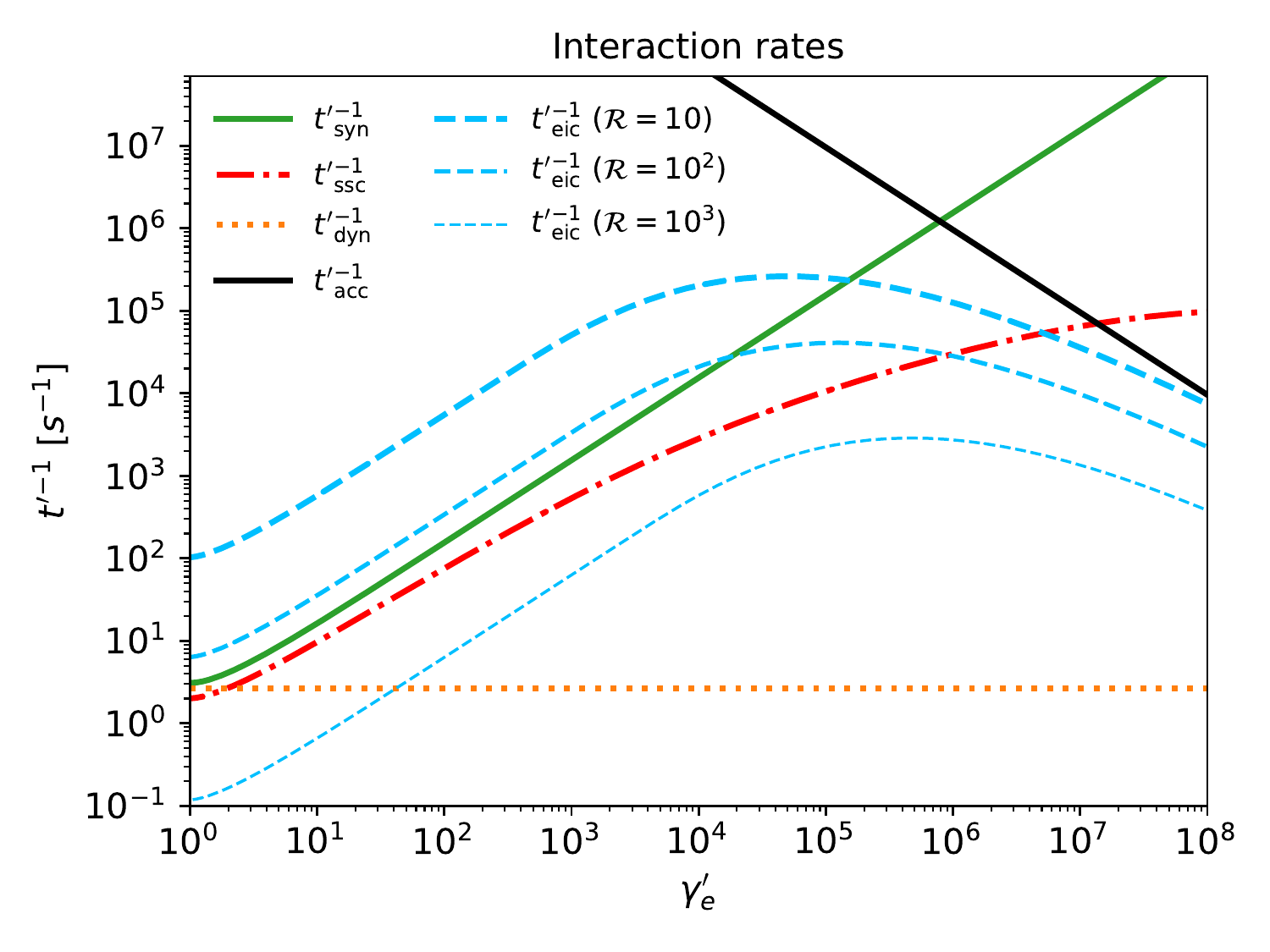}
\includegraphics[width=0.49\textwidth]{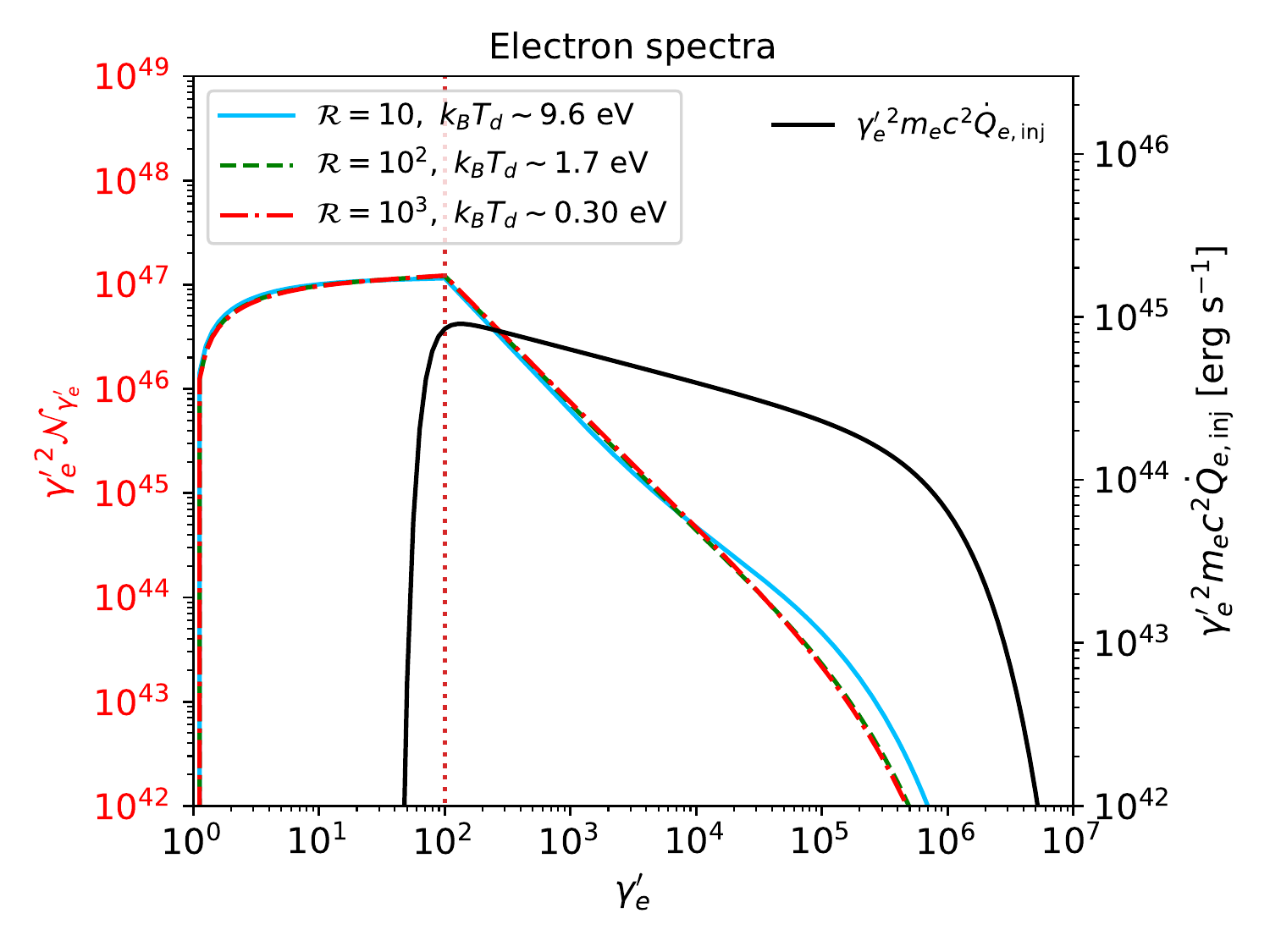}
\caption{\emph{Left panel:} Energy loss rates of accelerated electrons in the internal dissipation region. The green solid and red dash-dotted lines respectively show the synchrotron and SSC rates. From thick to thin, the blue dashed lines depict the EIC cooling rate for the CBOs at $\mathcal R=10, 10^2$ and $10^3$, respectively. The reciprocals of the dynamic and acceleration times are illustrated as the yellow dotted and black solid lines. \emph{Right panel:} The electron number spectra as functions of the electron Lorentz factor. The minimum injected Lorentz factor is $\gamma_{e,m}'=100$. The blue solid, green dashed and red dash-dotted lines correspond to $\mathcal R=10,~10^2$ and $10^3$ cases. The black solid line is the electron injection function.}
\label{fig:timescales}
\end{figure*}

\subsection{Disk photon spectra}
The accretion disk can become optically thick to ultraviolet/infrared photons as the plasma gets ionized. We estimate the vertical optical depth, for a fully ionized disk with temperature $T_d\gtrsim10^4~\rm K$, 
\begin{linenomath*}\begin{equation}\begin{split}
    \tau_{d}&\approx\Sigma_{\rm AGN}\kappa_{\rm R}\\
    &\simeq7.2\times10^3~(1+X)\dot m_\star M_{\star,8}\mathcal R_{2}^{-1/2}\alpha_{-1}^{-1}h_{\rm AGN,-2}^{-2},
    \end{split}
    \label{eq:tau_d}
\end{equation}\end{linenomath*}
where $\kappa_{\rm R}\approx0.2(1+X)$ is the Rosseland mean opacity for Thomson scattering and $X$ is the hydrogen mass fraction. Since the disk remains optically thick ($\tau_d>1$) in the range $\mathcal R\sim10-10^3$, we use a black-body spectrum to approximate the local photon density (in the units of $\rm eV^{-1}~cm^{-3}$), e.g., 
\begin{linenomath*}\begin{equation}
n_{\varepsilon_\gamma}^{(\rm eic)}=\frac{8\pi}{(hc)^3}\frac{\varepsilon_\gamma^2}{\exp\left(\frac{\varepsilon_\gamma}{k_BT_d}\right)-1},
\end{equation}\end{linenomath*}
{where $\varepsilon_\gamma$ is the energy of seed disk photons in the engine frame.}
The disk temperature $T_d$ at the position of the CBO can be written as \citep{2002apa..book.....F}
\begin{linenomath*}\begin{equation}\begin{split}
T_d&=\Bigg\{\frac{2GM_\star \dot M_\star}{8\pi\sigma_S R_d^3} \left[1-\left(\frac{R_{\rm *}}{R_d}\right)^{1/2}\right]\Bigg\}^{1/4}\\
&\simeq2.0\times10^4~\dot m_\star^{1/4}M_{\star,8}^{-1/4}\mathcal R_2^{-3/4}~{\rm K},
\end{split}\label{eq:temp_disk}
\end{equation}\end{linenomath*}
where $\sigma_S$ is the Stefan-Boltzmann constants and $R_{*}$ is the innermost edge of the disk. In this paper, we consider three distances $\mathcal R=10,~10^2$ and $10^3$. The corresponding disk temperatures are $k_BT_d=9.1~{\rm eV},~1.7{~\rm eV}$ and $0.3$ eV. For $R_d\gg R_{*}$, we have $T_d\propto \mathcal R^{-3/4}$, implying that the EIC component becomes increasing important when we move the CBO close to the central SMBH.

\section{Non-thermal electrons}\label{sec:electrons}
We consider a successful (i.e.non-choked) GRB jet {whose extended emission has a} luminosity $L_{j,\rm iso}=10^{48.5}~\rm{erg~s^{-1}}$. We focus on the internal dissipation model in which the jet kinetic energy is dissipated at $R_{\rm dis}=2\Gamma_j^2ct_{\rm var}\simeq1.5\times10^{12}~\Gamma_{j,1.7}^2t_{\rm var,-2}~{\rm cm}$ via internal shocks \citep{1994ApJ...430L..93R} or magnetic reconnections \citep{2012MNRAS.419..573M}, where $\Gamma_{j}=50\Gamma_{j,1.7}$ is the jet Lorentz factor, $t_{\rm var}=10^{-2}t_{\rm var,-2}~\rm s$ is the variability time of velocity fluctuations. One necessary condition for electron acceleration is that the upstream region should be optically thin for the shock not to be radiation mediated, namely, $\tau_{\rm in}=n'\sigma_TR_{\rm dis}/\Gamma_j\lesssim1$ \citep[e.g.,][]{2013PhRvL.111l1102M,2018PhRvD..98d3020K,2020PhRvD.102h3013Y}, where $n'=L_{j,\rm iso}/(4\pi R_{\rm dis}^2\Gamma_j^2m_pc^3)\simeq9.6\times10^{11}~L_{j,\rm iso,48.5}\Gamma_{j,1.7}^{-6}t_{\rm var}^{-1}{~\rm cm^{-3}}$ is the comoving number density and $\sigma_T$ is the Thomson cross section. Explicitly, we write down the optical depth as $\tau_{\rm in}\simeq1.8\times10^{-2}~L_{j,\rm iso,48.5}\Gamma_{j,1.7}^{-5}t_{\rm var,-2}^{-1}$, which indicates that efficient electron acceleration is plausible. 

To get the electron distribution we numerically solve the steady-state transport equation
\begin{linenomath*}\begin{equation}
    \frac{\mathcal N_{\gamma_e'}}{t_{\rm dyn}'}-\frac{\partial}{\partial \gamma_e'}\left(\frac{\gamma_e'}{t_{e,c}'}\mathcal N_{\gamma_e'}\right)=\dot Q'_{e,\rm inj},
    \label{eq:elec_dis}
\end{equation}\end{linenomath*}
where $\gamma_e'$ is the Lorentz factor, $\mathcal N_{\gamma_e'}=dN_e/d\gamma_e'$ is the differential spectrum, $t_{\rm dyn}'=R_{\rm dis}/(\Gamma_jc)$ is the dynamical time {that may represent adiabatic losses or escape}, $t_{e,c}'$ represents the electron cooling time scale, and the function $\dot Q_{e,\rm inj}$ is the electron injection rate from shock acceleration. Specifying a spectral index $s=2.2$, e.g., $\dot Q'_{e,\rm inj}\propto{\gamma_e'}^{-s}$, we normalize the injection function via $\int d\gamma_e'(\gamma_e'm_ec^2\dot Q'_{e,\rm inj})=\epsilon_eL_{j,\rm iso}/\Gamma_j^2$. The factor $\epsilon_e$, defined as the fraction of jet kinetic energy that is converted to electrons, is assumed to be $\epsilon_e=0.1$. The minimum Lorentz factor $\gamma_{e,m}'$ for injected electrons is assumed to be $\gamma_{e,m}'=100$.  

In the dissipation region, the magnetic field is $B'_{\rm dis}=[8\pi\epsilon_B(\Gamma_{\rm rel}-1)n'm_pc^2]^{1/2}\simeq3.8\times10^4~\epsilon_{B,-2}^{1/2}L_{j,\rm iso,48.5}^{1/2}\Gamma_{j,1.7}^{-3}t_{\rm var}^{-1/2}~{\rm G}$, where $\Gamma_{\rm rel}\simeq5$ is the relative Lorentz factor between the fast and slow shells. {The ratio of $B'_{\rm dis}$ to the disk magnetic field $B_d$ is $B_{\rm dis}'/(\Gamma_jB_{d})\simeq3.8~\mathcal R_{2}^{9/8}\beta_{0.48}^{1/2}$. Here, we focus on the $\mathcal R$-dependence of the magnetic fields, using the fiducial values for all other parameters. We use the modulated magnetic field $B'=\max[B_{\rm dis}',\Gamma_jB_d]$ to calculate the electromagnetic emission in the dissipation region.} 

The accelerated electrons lose energy through synchrotron, SSC, and EIC processes within the corresponding timescales $t_{e,\rm syn}'$, $t_{e,\rm ssc}'$ and $t_{e,\rm eic}'$. The net cooling timescale is given by $t_{e,c}'=({t'}^{-1}_{e,\rm syn}+{t'}^{-1}_{e,\rm ssc}+{t'}^{-1}_{e,\rm eic})^{-1}$. Electrons with higher $\gamma_e'$ cool down faster while a longer acceleration time, e.g., $t'_{\rm acc}=\gamma'_{e}m_ec/(eB')$, is required to reach such a high energy. We thus expect a cutoff Lorentz factor $\gamma'_{e,\rm cut}$ determined by the equation $t'_{\rm acc}=t'_{e,c}$, above which electrons cannot accumulate energy due to the rapid radiation. Using these arguments, the injection function for a spectral index $s>2.0$ can be written as,
\begin{linenomath*}\begin{equation}
\dot Q'_{e,\rm inj}=\frac{(s-2)\epsilon_eL_{j,\rm iso}}{\Gamma_j^2m_ec^2\gamma{'}^2_{e,m}}\left(\frac{\gamma'_e}{\gamma{'}_{e,m}}\right)^{-s}\exp\left(-\frac{\gamma'_e}{\gamma'_{e,\rm cut}}\right).
\label{eq:e_inj}
\end{equation}\end{linenomath*}
The photons from the synchrotron process play the role of seed photons in EIC scattering. Therefore, we need a trial electron spectrum, e.g. $\mathcal N_{\gamma'_e}^{(\rm 0)}\sim(t{'}^{-1}_{\rm dyn}+t{'}^{-1}_{e,\rm syn}+t{'}^{-1}_{e,\rm eic})^{-1}\dot Q'_{e,\rm inj}$, to evaluate $t'_{\rm ssc}$, and solve the differential equation \ref{eq:elec_dis} iteratively to obtain a convergent solution as in \cite{2021ApJ...920...55Z}.

The left panel in Fig. \ref{fig:timescales} shows the energy loss rates. The blue dashed lines show the EIC cooling rate for $\mathcal R=10,~10^2$ and $10^3$. The synchrotron (green line) and SSC (red dash-dotted line) cooling rates are not sensitive to the CBO's position, whereas the EIC rate increases as the distance between the CBO and the SMBH reduces. This tendency is consistent with equation \ref{eq:temp_disk}, which predicts a hotter and photon-denser environment close to the SMBH. Remarkably, the EIC process starts to dominate the electron cooling at a distance range $\mathcal R\lesssim10^2$, leading to a softer electron spectrum, e.g., the blue line ($\mathcal R=10$) in the right panel of Fig. \ref{fig:timescales}, in contrast to the high-$\mathcal R$ cases. The black solid line in the right panel shows the electron injection function. In the low-energy band, there is no injection, e.g., $\dot Q_{e,\rm inj}=0$ for $\gamma_e'\lesssim\gamma'_{e,m}$, we can analytically solve equation \ref{eq:elec_dis} and connect this segment to the $\gamma_e'>\gamma'_{e,m}$ part. Using the simplification $t{'}_c^{-1}\sim b\gamma'_{e}$, which is consistent with the EIC and synchrotron cooling rates in the left panel, we obtain
\begin{linenomath*}\begin{equation}
    \mathcal N_{\gamma_e'}=\mathcal N_{\gamma_{e,m}'}\exp\left[-\frac{1}{bt'_{\rm dyn}}(\gamma'_{e,m}-\gamma'_e)\right],~\gamma_e'\lesssim\gamma'_{e,m},
    \label{eq:lowE_e}
\end{equation}\end{linenomath*}
where $\mathcal N_{\gamma_{e,m}'}$ represents the electron number distribution at $\gamma_{e,m}'$.
Equation \ref{eq:lowE_e} explains the electron spectrum softening at {lower values of $\mathcal R$ (equivalently at larger values of $b$).}

\begin{figure}\centering
\includegraphics[width=0.49\textwidth]{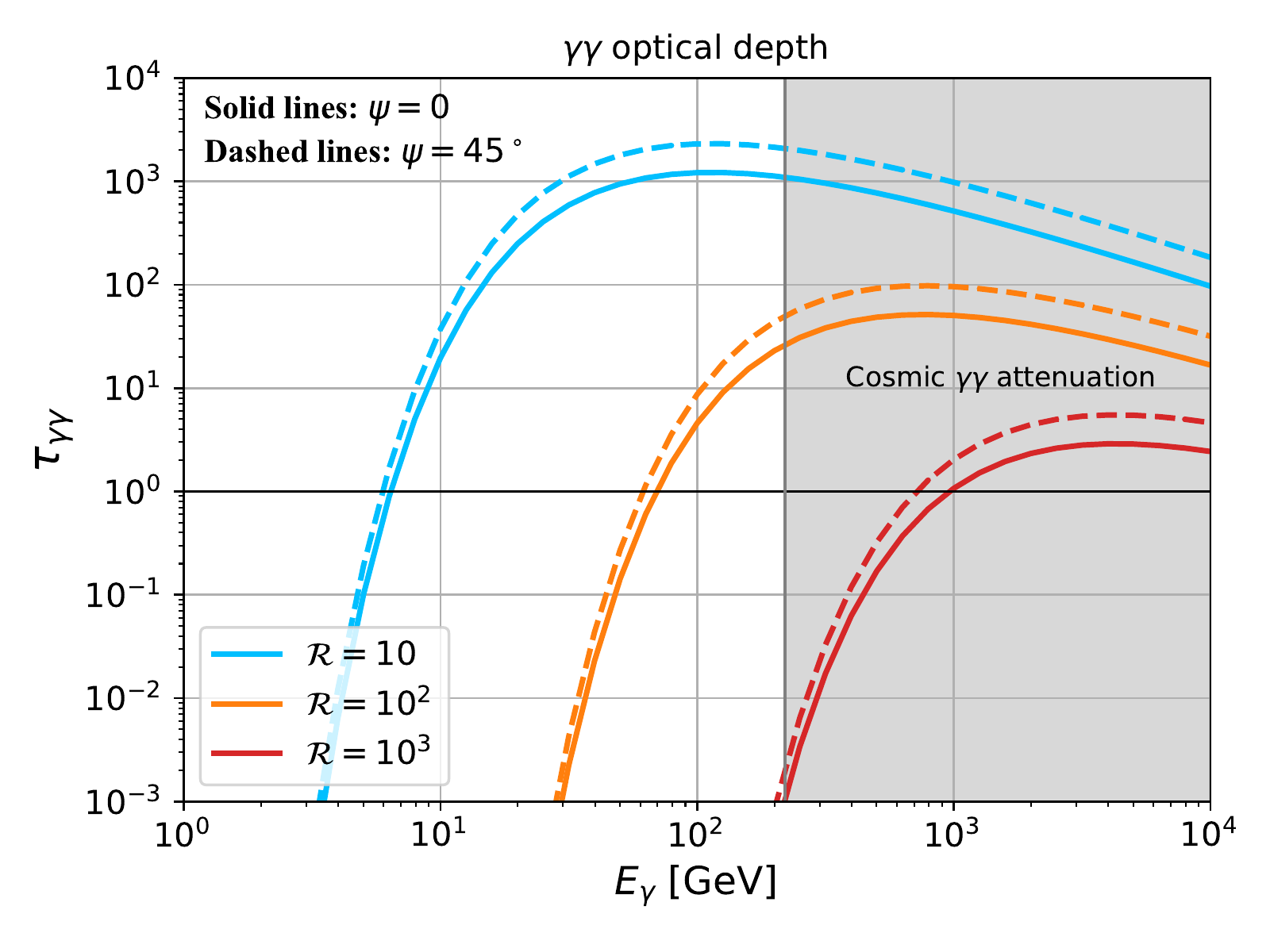}
\caption{The blue ($\mathcal R=10$), yellow ($\mathcal R=10^2$) and red ($\mathcal R=10^3$) lines are the optical depth $\tau_{\gamma\gamma}$ for $\gamma\gamma$ annihilation between $\gamma$-rays and disk photons. The solid and dashed lines correspond to the inclination $\psi=0$ and $\psi=45^\circ$. The optical depth to cosmic $\gamma\gamma$ annihilation becomes greater than $1.0$ in the energy range $E_\gamma\gtrsim220~\rm GeV$ (the gray shaded area), assuming that the CBO merger is located at $z=1.0$.}
\label{fig:tau_gg}
\end{figure}

\begin{figure*}\centering
\includegraphics[width=0.99\textwidth]{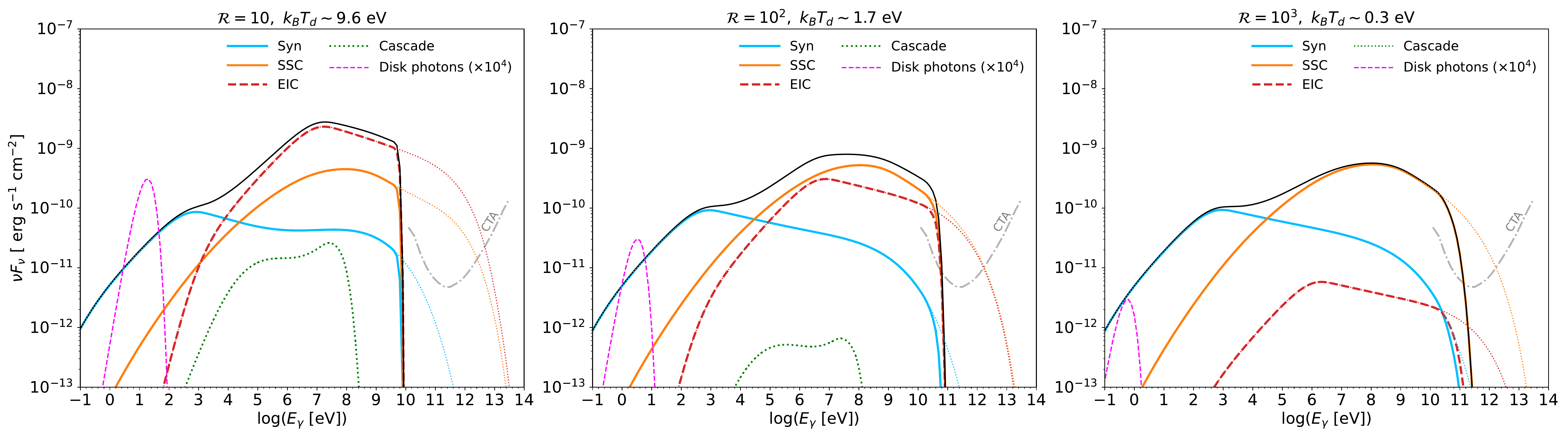}
\caption{The observed $\gamma$-ray spectra from embedded short GRBs at $z=1$ with distances $\mathcal R=10$ (left panel), $10^2$ (middle panel) and $10^3$ (right panel) to the central SMBH. The GRB parameters used here are the fiducial parameters assumed in \S\ref{sec:electrons}, e.g., $L_{j,\rm iso}=10^{48.5}~\rm erg~s^{-1}$, $\Gamma_{j}=50$, $\epsilon_B=0.01$, and $\epsilon_e=0.1$. The blue, yellow and red solid lines show the synchrotron, SSC and EIC emission after $\gamma\gamma$ attenuation. {The dotted lines in the corresponding colors depict the unattenuated fluxes.} The cascade emissions are depicted as the green lines. {The magenta dashed lines show the disk target photon fluxes (multiplied by $10^4$).} In both cases, $\psi=0$ is applied. {The gray dash-dotted lines indicate the CTA flux sensitivity for the $10^3~\rm s$ observation time.}}
\label{fig:EMspectra}
\end{figure*}
 
 
\section{Results}\label{sec:GAMMAspectra} 
\subsection{$\gamma$-ray spectra}
Using the electron spectra obtained in \S\ref{sec:electrons} and following the formalism and procedures presented in \cite{2011ApJ...732...77M}, \cite{2021ApJ...920...55Z} and \cite{2021ApJ...911L..15Y}, we numerically compute the $\gamma$-ray spectra taking into account the synchrotron, SSC and EIC processes. We consider three merger-induced GRBs in an AGN located at redshift $z=1$ (the equivalent luminosity distance is $d_L\simeq6.7~\rm Gpc$). We focus on the on-axis case and assume the CBOs' orbit planes are all aligned with the AGN disk {plane}, e.g., $\psi=0$. A discussion on the influence of $\psi$ will be given in \S\ref{sec:detection}. 

While propagating in the jet and in the AGN disk, high-energy $\gamma$-rays will annihilate with ambient UV/IR disk photons, resulting in their attenuation and EM cascades. The optical depth for $\gamma\gamma$ annihilation depends on the photon energy in the short GRB's engine frame  $\varepsilon_\gamma=\Gamma_j\varepsilon_{\gamma}'$, the position of the jet and the misalignment angle $\psi$, via
\begin{linenomath*}\begin{equation}
    \tau_{\gamma\gamma}[\varepsilon_\gamma,\mathcal R,\psi]\approx\int_0^{H_{\rm AGN}}\frac{dy}{\cos\psi}{\lambda^{-1}_{\gamma\gamma}[\varepsilon_\gamma,R_d+y\tan\psi]},
\end{equation}\end{linenomath*}
where the reciprocal of the mean free path $\lambda_{\gamma\gamma}[R_d]$ for an isotropic disk photon field can be calculated as \citep[e.g.,][]{2009herb.book.....D}
\begin{linenomath*}\begin{equation}
    \lambda^{-1}_{\gamma\gamma}[\varepsilon_\gamma,R_d]=\frac{1}{2}\int_{-1}^1d\mu(1-\mu)\int d\tilde\varepsilon_\gamma n_{\epsilon_\gamma}^{\rm(eic)}[\tilde\varepsilon_\gamma]\sigma_{\gamma\gamma}[x].
\end{equation}\end{linenomath*}
In this expression, $x=\tilde\varepsilon_\gamma\varepsilon_\gamma(1-\mu)/2$ is the particle Lorentz factor in the center-of-momentum frame and $\sigma_{\gamma\gamma}$ is the $\gamma\gamma$ annihilation cross section. 
 
Fig. \ref{fig:tau_gg} shows the optical depth in the observer's frame, where the observed energy is connected with $\varepsilon_\gamma$ and $\varepsilon_{\gamma}'$ via $E_\gamma=\varepsilon_\gamma/(1+z)=\Gamma_j\varepsilon_\gamma'/(1+z)$. The solid blue, yellow and red lines illustrates $\tau_{\gamma\gamma}$ at $\mathcal R=10,~10^2$ and $10^3$ with $\psi=0$, whereas the dashed lines correspond to the case of an inclined jet, e.g., $\psi=45^\circ$. The universe becomes opaque for $\gamma$-rays produced at $z=1$ with energies $E_{\gamma}\gtrsim220~\rm GeV$ (see the gray area in Fig. \ref{fig:tau_gg}) due to $\gamma\gamma$ annihilation between $\gamma$-rays and cosmic backgrounds \citep{2010ApJ...712..238F} e.g. extragalactic background light (EBL) and cosmic microwave background (CMB).  From Fig. \ref{fig:tau_gg}, we find that $\gamma$-rays with energy $E_{\gamma}\gtrsim10~\rm GeV$ are strongly suppressed due to $\gamma\gamma$ annihilation for a GRB close to the SMBH, i.e. $\mathcal R\simeq10$. For a GRB at positions with a larger $\mathcal R\sim10^2-10^3$, $\gamma$-ray photons with energy $E_\gamma\sim100~\rm GeV$ can escape from the AGN disk.

Applying the factor $\exp(-\tau_{\gamma\gamma})$ to the $\gamma$-ray spectra, we obtain the $\gamma\gamma$-attenuated spectra for embedded GRBs at redshift $z=1$, as shown in Fig. \ref{fig:EMspectra}. In this figure, $\psi=0$ is used. The blue solid, yellow solid, and red dashed lines respectively illustrate the synchrotron, SSC, and EIC components. The dotted lines with corresponding colors show the fluxes before $\gamma\gamma$ attenuation. {The gray dash-dotted lines indicate the Cherenkov Telescope Array (CTA) flux sensitivity for the $10^3$ s observation time \citep{al2019science}.} {The magenta dashed lines show the disk photon fluxes multiplied by $10^4$.} From the red dashed lines in Fig.  \ref{fig:EMspectra}, we find that a closely embedded GRB can produce brighter $\gamma$-ray emission due to the EIC enhancement. {The ``Compton dominance" induced by EIC enhancement can be used as the prominent feature to distinguish these embedded short GRBs from others.}
 
The $e^{+}/e^{-}$ pairs produced in the $\gamma\gamma$ annihilation process will induce electromagnetic cascades while diffusing and cooling down in the AGN disk via synchrotron and inverse Compton processes. {Following the treatment in
\cite{2007ApJ...671.1886M}, we write down the distribution for the secondary electrons and positrons,
\begin{linenomath*}\begin{equation}
\mathcal N_{\gamma_e}^{\rm cas}\approx 2\mathcal N_{\hat\varepsilon_\gamma}^{\rm ph}\left(\frac{d\hat\varepsilon_\gamma}{d\gamma_e}\right)\left(1-e^{-\tau_{\gamma\gamma[\hat\varepsilon_\gamma,\mathcal R,\psi]}}\right)
\end{equation}\end{linenomath*}
where $N_{\hat\varepsilon_\gamma}^{\rm ph}$ is the pre-attenuation gamma-ray number spectra (in the units of $\rm eV^{-1}$) in the engine frame and $\hat\varepsilon_\gamma=2\gamma_em_ec^2$ is the energy of primary electrons.} {Using the cavity magnetic field $B_{\rm cav}\approx(2\epsilon_B\eta_w\dot M_{\rm CBO}v_w/H_{\rm AGN}^2)^{1/2}\simeq98~\epsilon_{B,-2}^{1/2}\eta_{w,-0.5}^{1/2}h_{\rm AGN.-2}^{-1}\eta_{\rm CBO,-1}^{1/2}\mathcal R_{2}^{-1}\dot m_{\star}^{1/2}M_{\star,8}^{-1/2}v_{w,9}^{1/2}~{\rm G}$,} we numerically calculate the cascade emission. The green dotted lines in Fig. \ref{fig:EMspectra} show the cascade emission. Comparing to the beamed emission produced in the jet, the cascade emission is subdominant for $\mathcal R\gtrsim 100$ and typically peaks at a lower energy $\sim100~\rm MeV$. We find that the cascade flux drops dramatically as $\mathcal R$ increases, which is consistent with the $\mathcal R$-dependence of the $\gamma\gamma$ optical depth in Fig. \ref{fig:tau_gg}. When the disk becomes transparent to the $\gamma$-ray photons, the $e^-/e^+$ pair production is suspended and the cascade emission is strongly suppressed. {Typically, we need to solve the time-dependent equations to obtain the secondary electron/position distributions and the cascade spectrum. Our approach can provide a good estimation since these secondary particles cools down very fast, e.g., $t_{e, c}^{\rm cas}\lesssim10~\rm s.$}

\begin{figure*}
    \centering
    \includegraphics[width=0.49\textwidth]{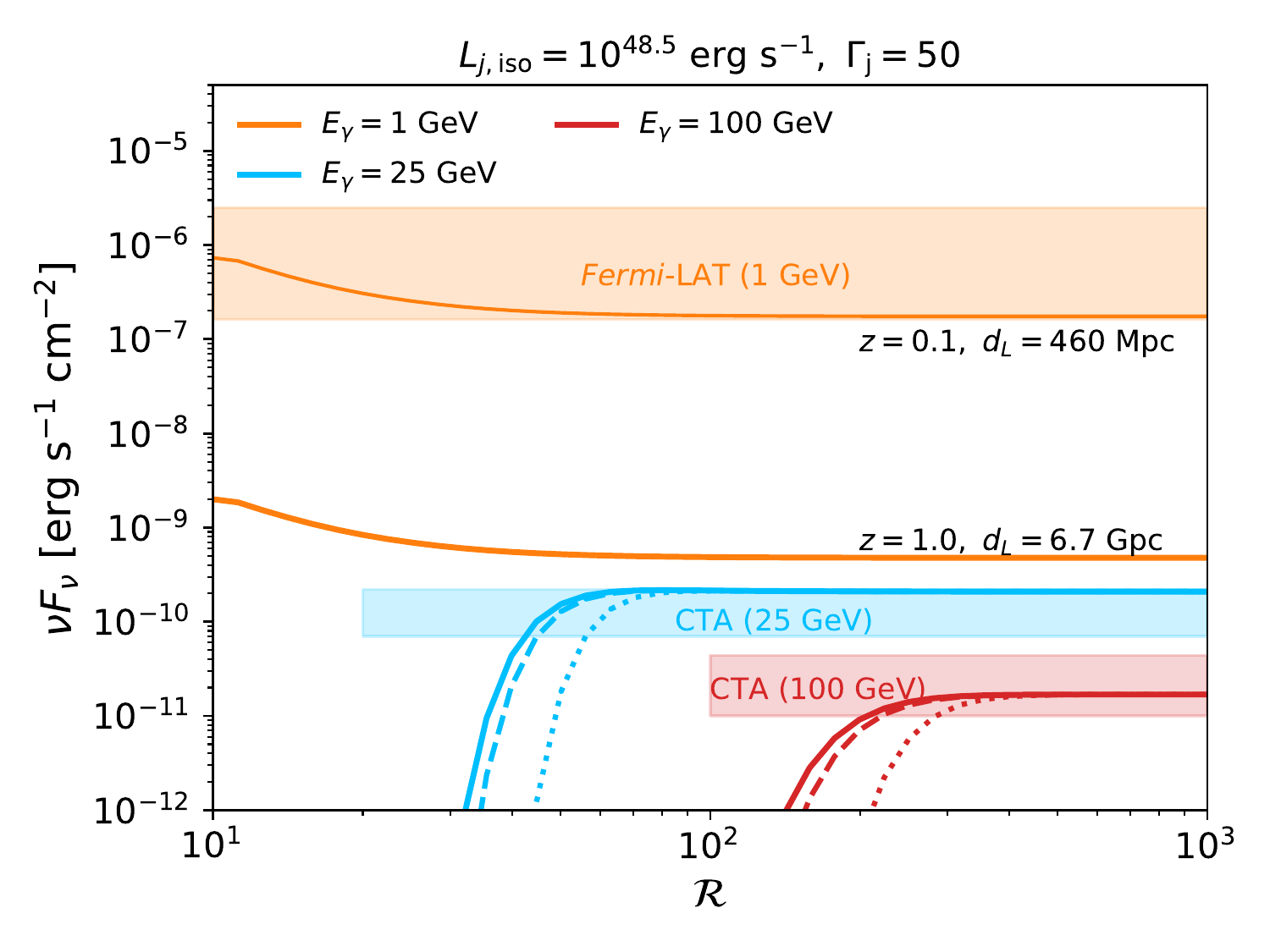}
    \includegraphics[width=0.49\textwidth]{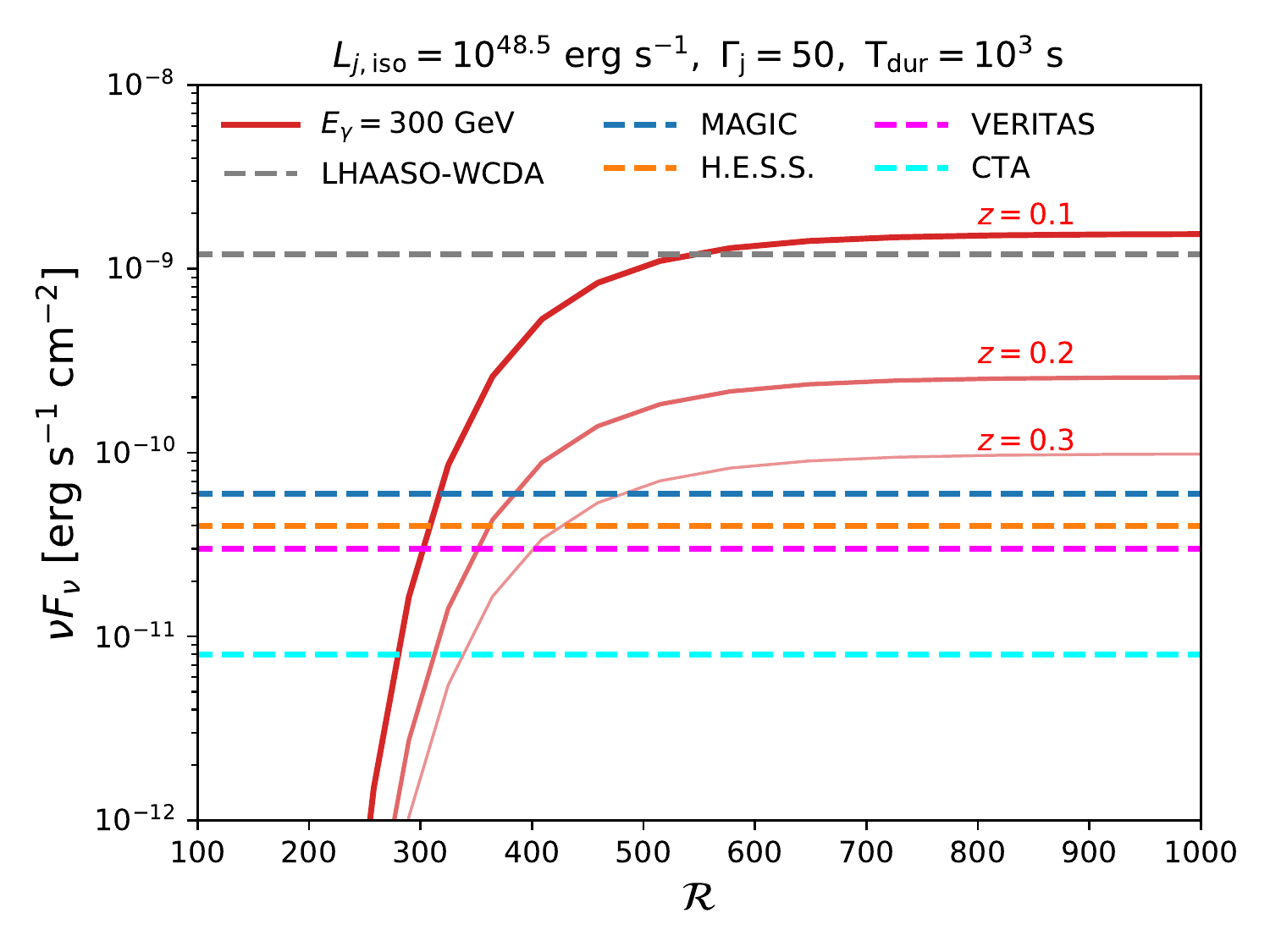}
    \caption{\emph{Left panel:} $\gamma$-ray fluxes at 1 GeV (yellow lines), 25 GeV (blue lines) and 100 GeV (red lines) as functions of $\mathcal R$. The thick lines are obtained with $L_{j,\rm iso}=10^{48.5}~\rm erg~s^{-1}$ and $z=1.0$, whereas a closer short GRB at $z=0.1$ is considered for the thin yellow line. The point-source performance for \emph{Fermi}-LAT and CTA at corresponding energies are shown as the yellow, blue and red areas, respectively. The upper and lower bounds show the sensitivities for the observation time $T_{\rm dur}=10^2$ s and $T_{\rm dur}=10^3$ s.
    {\emph{Right panel:} The red solid lines from thick to thin show the $\mathcal R$-dependence of 300 GeV $\gamma$-ray fluxes from the embedded short GRBs at $z=0.1$, 0.2, and 0.3. The horizontal dashed lines from top to bottom correspond the the sensitivities of LHAASO-WCDA, MAGIC, H.E.S.S., VERITAS, and CTA.}
    }
    \label{fig:fluxRd}
\end{figure*}

\subsection{Detectability with Fermi-LAT and VHE $\gamma$-ray facilities}
\label{sec:detection}
It is useful to compare the expected $\gamma$-ray fluxes {in the extended emission phase} {against} the sensitivities of current and future facilities, {such as \emph{Fermi}-LAT, MAGIC, H.E.S.S., VERITAS, CTA, and LHAASO-WCDA}, and discuss how the parameters $\mathcal R$ and $\psi$ influence the results. 

{Observationally, a significant fraction of short GRBs exhibit `long-lasting' extended or plateau emission peaking in X-ray bands \citep{2006ApJ...643..266N,2011ApJS..195....2S,2015MNRAS.452..824K,2017ApJ...846..142K} with the duration $T_{\rm dur}\sim10^2-10^5$ s, following the prompt phase where $~90\%$ of the kinetic energy is dissipated in $\sim2$ seconds, e.g., $T_{90}\lesssim2~\rm s$. Such prolonged emission may originate from the continuous energy injection by the accreting black holes formed after the merger or the fast rotating magnetars \citep[e.g.,][]{2006Sci...311.1127D,2008MNRAS.385.1455M,2011MNRAS.417.2161B,2012MNRAS.419.1537B,2013MNRAS.430.1061R,2014MNRAS.438..240G,2015ApJ...804L..16K}.}
Considering a prolonged $\gamma$-ray emission of luminosity $L_{j,\rm iso}=10^{48.5}~\rm erg~s^{-1}$ and the corresponding duration in the observer's frame $T_{\rm dur}\sim10^2{~\rm s}-10^{3}~{\rm s}$, we show the integral sensitivities within $T_{\rm dur}$ for \emph{Fermi}-LAT\footnote{The \emph{Fermi}-LAT sensitivity can be found in \url{https://www.slac.stanford.edu/exp/glast/groups/canda/lat_Performance.htm}} and CTA \citep{al2019science} at the $E_\gamma=$ 1 GeV (yellow area), 25 GeV (blue area) and 100 GeV (red area) in the left panel of Fig. \ref{fig:fluxRd}. The upper and lower bounds of each shaded area demonstrate the performances for the detectors given the observation time $T_{\rm dur}=10^2~\rm s$ and $T_{\rm dur}=10^3~\rm s$, respectively. We plot also the $1~\rm GeV$ (yellow lines), $25~\rm GeV$ (blue lines) and $100~\rm GeV$ (red lines) fluxes as functions of $\mathcal R$ in the left panel of Fig. \ref{fig:fluxRd}. The solid lines correspond to the $\psi=0$ case, whereas the dashed and dash-dotted lines depict the $\psi=45^\circ$ and $\psi=75^\circ$ cases. The thick lines are for the GRBs at $z=1$, while the thin yellow line shows the 1 GeV fluxes for a closer GRB at $z=0.1$ ($d_L\simeq460~\rm Mpc$). 

The influence of disk photons is encoded in the shapes of the yellow, blue and red curves. The $1$ GeV flux decreases to a flat level as $\mathcal R$ increases because the EIC component gradually becomes less important as the CBO is moved to a cooler outer region. In the ranges $\mathcal R\lesssim50$ and $\mathcal R\lesssim300$, the $\gamma\gamma$ attenuation caused by dense disk photons suppresses the $25$ GeV and $100$ GeV emission, respectively. Since the $\gamma\gamma$ annihilation is negligible for 1 GeV photons even if the CBO is very close to the SMBH (see the blue lines in Fig. \ref{fig:tau_gg}), we expect that the flux does not depend on $\psi$. On the other hand, the $25$ GeV and 100 GeV fluxes decrease as $\psi$ approaches $\psi_c\simeq85.6^\circ.$

From the left panel of Fig. \ref{fig:fluxRd}, we find that CTA will be capable of detecting 25 GeV and 100 GeV $\gamma$-rays up to $z=1$ if an embedded short GRB is appropriately distant from the SMBH, e.g., $\mathcal R\gtrsim40$ for 25 GeV $\gamma$-rays and $\mathcal R\gtrsim200$ for 100 GeV $\gamma$-rays. By contrast, it is challenging for \emph{Fermi}-LAT to detect the 1 GeV photons from sources located at $z=1$ via point source search within the duration $T_{\rm dur}\sim10^3$ s. For the short GRBs embedded in AGN disks, we would require a nearby CBO merger ($d_L\lesssim460\rm~Mpc$) at the position with the distance greater than $40R_{\rm S}$ ($\mathcal R\gtrsim40$) to the central SMBH in order to be detected simultaneously by CTA and \emph{Fermi}-LAT. 

{MAGIC, H.E.S.S., and VERITAS are current ground Imaging Atmospheric Cherenkov Telescopes with very good performance in the energy range 150 GeV to 30 TeV. LHAASO is a new generation multi-component instrument and LHAASO-WCDA is operated in the energy range $\sim$300 GeV to 10 TeV. We present the $\mathcal R$-dependence of 300 GeV $\gamma$-ray fluxes at $z=0.1$, 0.2, and 0.3 (the red solid lines, from thick to thin) in the right panel of Fig. \ref{fig:fluxRd}. The horizontal dashed lines from top to bottom corresponds to the flux sensitivities of LHAASO-WCDA \citep{2019arXiv190502773B}, MAGIC  \citep{2016APh....72...76A}, H.E.S.S. \citep{2015arXiv150902902H}, VERITAS\footnote{The differential sensitivity of VERITAS can be found in \url{https://veritas.sao.arizona.edu/about-veritas/veritas-specifications}}, and CTA for $T_{\rm dur}=10^3~\rm s$ and $\psi=0$. 
At 300 GeV, the sensitivity of LHAASO-WCDA is $\sim10^{-9}~\rm erg~s^{-1}~cm^{-2}$ in $10^3$ s observation. The nearby embedded GRBs with redshift $z<0.1$ can be observed.
MAGIC, H.E.S.S., VERITAS and CTA can detect 300 GeV photons from embedded GRBs upto redshift $z=0.3$ if $\mathcal R\gtrsim500$ is satisfied. For the sources with farther distance, the Universe could be opaque to VHE $\gamma$-rays. }

\subsection{Prompt emission}
\label{sec:prompt}
{As for the prompt emission, besides the cutoff with energy $\gtrsim100$ GeV caused by the $\gamma\gamma$ absorption in the AGN disk, we found that there may be no significant difference between short GRBs embedded in AGN disks and other short GRBs. The reason is that, given a higher isotropic luminosity $L_{j,\rm iso}^{\rm prompt}=10^{51}~\rm erg~s^{-1}$ and a higher Lorentz factor $\Gamma_{j}^{\rm prompt}=200$ ($\Gamma_{j}^{\rm prompt}=100$) in the prompt emission phase of $T_{90}=1$ s, the EIC emission is subdominant (comparable) compared to the synchrotron/SSC components. Using the parameters in the prompt emission phase, we estimate photon flux in the energy range 50 - 300 keV,
\begin{linenomath*}\begin{equation}
F_{\nu,50-300~\rm keV}^{\rm prompt}\simeq 1.9~(1+z)d_{L,28}^{-2} ~\rm ph~s^{-1}~cm^{-2}.
\end{equation}\end{linenomath*}
Noting that the onboard trigger threshold of the \emph{Fermi} Gamma-Ray Burst Monitor (\emph{Fermi}-GBM) is $\sim0.7~\rm ph~s^{-1}~cm^{-2}$ \citep{2009ApJ...702..791M}, it can detect the prompt emission and localize the short GRB. At 10 GeV, the flux of the prompt emission is $\nu F_{\nu,\rm 10~GeV}^{\rm prompt}\sim2\times10^{-6}~(1+z)d_{L,28}^{-2}~\rm erg~{s^{-1}}~cm^{-2}$, implying the possible detection of 
the embedded GRBs at $z\sim0.5-1$ with the High Altitude Water Cherenkov (HAWC) observatory \citep{2012APh....35..641A}. If the short GRB is GRB 090510-like, e.g., $L_{j,\rm iso}^{\rm prompt}\gtrsim10^{53}~\rm erg~s^{-1}$, \emph{Fermi}-LAT would also be able to see $\gamma$-ray photons upto $\sim30$ GeV in the prompt emission phase \citep{2010ApJ...716.1178A}. Above all, the prompt emission diagnosis can provide valuable information for the follow-up observations of extended emissions.} 

\section{Summary and discussion}\label{sec:summary}
We studied $\gamma$-ray emission from short GRBs embedded in AGN disks and showed that successful jets are expected from these, since the CBOs in the disks are highly super-Eddington accretors and can produce low-density cavities around the CBO via powerful outflows. Our work demonstrates that the AGN disks influence the $\gamma$-ray emission mainly in two ways, namely, via the EIC enhancement and $\gamma\gamma$ attenuation, depending on the distance to the SMBH and the inclination $\psi$. If a CBO merger occurs very close to the SMBH, e.g., $\mathcal R\sim10-40$, the dense disk photon field will lead to a luminous EIC component in the GeV band and a firm cutoff at $E_\gamma\simeq 10$ GeV. On the other hand, the SSC process dominates the GeV emission for CBO mergers at $\mathcal R\gtrsim100$, and the disk gradually becomes transparent for 10-100 GeV photons unless the GRB jet is entirely buried inside the AGN disk, e.g., $\psi\gtrsim\psi_c\simeq85.6^\circ$. {Considering the ratio of the peak flux of the inverse Compton component to the synchrotron peak flux and the cutoff energy, we may be able to distinguish the short GRBs embedded in AGN disks from other types of isolated short GRBs \citep[e.g.,][]{2018ApJ...854...60M,2019ApJ...887L..16K}}. {To identify the embedded short GRBs, we can utilize these two signatures, ``Compton dominance" and $\gamma\gamma$ annihilation cutoff. }
Such spectral information can also be used to determine the parameters of the short GRB - AGN disk system such as $T_d$, $\mathcal R$ and $\psi$. {According to the simulations of compact binary formations in AGN disks, it is reasonable to expect the embedded short GRBs to occur in the region $\mathcal R\gtrsim 40-100$ \citep{2016ApJ...819L..17B,2020ApJ...898...25T}. The detection of these short GRBs can, in return, be used to test current AGN-assisted CBO formation theories and constrain the CBO distributions in AGN disks.}

Since approximately $f_{\rm EE}\sim1/4-1/2$ \citep[e.g.,][]{2012MNRAS.419.1537B} of \emph{Swift} short GRBs are accompanied by extend emission, we investigated the detectability {of GRBs in the AGN disk} for CTA and \emph{Fermi}-LAT considering a jet of luminosity $L_{j,\rm iso}=10^{48.5}~\rm erg~s^{-1}$ lasting for $T_{\rm dur}\sim10^2-10^3$~s. From now on, we discuss the detection perspectives of the extended emissions with $T_{\rm dur}=10^2-10^3$ s, $L_{\rm j,\rm iso}=10^{48.5}~\rm erg~s^{-1}~cm^{-2}$, and $\Gamma_j=50$. For the embedded short GRBs within $z=1.0$, CTA will be able to detect the $\gamma$-rays in the energy range $E_\gamma\sim25-100$ GeV if the requirements $\mathcal R\gtrsim\mathcal R_c$ and $\psi\lesssim\psi_c$ are satisfied, where $\mathcal R_c\sim40-100$ is the critical distance defined by $\tau_{\gamma\gamma}[(1+z)E_\gamma,\mathcal R_c,\psi]=1$. To estimate the CTA detection rate, we use $f_{\mathcal R}$ and $f_{\psi}\sim1$ to represent the fractions of embedded short GRBs that meet the conditions $\mathcal R\gtrsim\mathcal R_c$ and $\psi\lesssim\psi_c$, respectively. Taking into account both NS-NS and NS-BH mergers, \cite{2020MNRAS.498.4088M} estimated the occurance rate of short GRB in AGN disks at $z<1$, $\dot R_{\rm SGRB,AGN}\sim(300-2\times10^4)f_{\rm AGN,-1}~\rm yr^{-1}$, where $f_{\rm AGN}\sim0.1$ is the fraction of BH-BH mergers. We estimate the CTA detection rate of the on-axis prolonged $\gamma$-ray emission from short GRBs embedded in AGN disks via $\dot R_{\rm CTA}\sim f_{\rm CTA}f_{b}f_{\rm EE}f_{\mathcal R}f_{\psi}\dot R_{\rm GRB,AGN}\sim(0.2-22)~f_{\mathcal R}\theta_{j,-1}^2f_{\rm AGN,-1}~\rm yr^{-1}$, where $f_{\rm CTA}\sim0.3-0.5$ is the CTA detection efficiency defined as the ratio of detectable events to events that can be followed up by CTA \citep[e.g.,][]{2013APh....43..252I}, $f_{b}=(\theta_j+1/\Gamma_j)^2/2\sim\theta_j^2/2$ is the beaming factor and $\theta_j\sim0.1$ is the jet opening angle. Despite the large uncertainty in the CTA detection rate, {we estimate that it is feasible for} CTA to detect the prolonged $\gamma$-ray emission from short GRBs embedded in AGN disks in the time scale of one year. 

We now discuss the implications to multi-messenger analyses with GWs and $\gamma$-rays. \cite{2017ApJ...835..165B} estimated that the merger rate of binary black holes (BBHs) embedded in AGN disks within the advanced Laser Interferometer Gravitational-wave Observatory's (aLIGO's) horizon, e.g., $D_h\simeq450~\rm Mpc$, could be $\dot R_{\rm L,BBH}\sim20~\rm yr^{-1}$. Implementing the ratio of the cumulative NS-BH and NS-NS merger rates to the BBH merger rate in the AGN channel, $f_{\rm L,CBO/BBH}=(\dot R_{\rm L,NS-NS}+\dot R_{\rm L,NS-BH})/\dot R_{\rm L,BBH}\sim0.1-7.0$ \citep{2020MNRAS.498.4088M}, we estimate the occurrence rate of on-axis short GRBs with extended emission originating from LIGO-detectable CBO mergers in the AGN channel,
\begin{linenomath*}\begin{equation}\begin{split}
    \dot R_{\rm SGRB-AGN}^{(L)}&=f_{\rm EE}f_bf_{\rm L,CBO/BBH}\dot R_{\rm L,BBH}\\
    &\sim(2.5\times10^{-3}-0.35)~\theta_{j,-1}^{2}~\rm yr^{-1}.
    \end{split}
\end{equation}\end{linenomath*}
The physical meaning of this equation is that among all detectable mergers within LIGO's horizon, {MAGIC, H.E.S.S., VERITAS, CTA, and LHAASO-WCDA} can observe $2.5\times10^{-3}-0.35$ short GRBs with extended $\gamma$-ray emission each year. In the optimistic case, it is possible to detect the on-axis extended emission simultaneously with GWs originated from CBO mergers embedded in AGN disks in one decade. 


{We note also that, while this is not the subject of the present work, the model predicts that short GRBs from CBO mergers are efficient neutrino emitters. Our model does not require choked jets, unlike \citet{2021ApJ...911L..19Z,2021ApJ...906L..11Z}.
The CRs accelerated in the successful jet can efficiently interact with disk photons and produce high-energy neutrinos via the photomeson production process. Using equations 8 and 9 of \citet{2016PhRvL.116g1101M} and Fig. \ref{fig:tau_gg} of this work, the photomeson optical depth is $f_{p\gamma}\sim1$ for ${\mathcal R}\sim10$ and $f_{p\gamma}\sim0.1$ for ${\mathcal R}\sim100$. High-energy neutrinos are expected in the PeV range, and they will make additional contribution to those predicted by \citet{2017ApJ...848L...4K}. The enhancement is more prominent for prompt neutrino emission, because the efficiency is low for usual short GRBs.}

{In conclusion, future multi-messenger analyses of AGN short GRBs can provide unprecedented insights for} understanding the  formation and evolution of CBOs inside the AGN disks as well as on the origin of their high-energy emission.

\section*{Acknowledgements}
We thank B. Theodore Zhang, Mukul Bhattacharya, Zsuzsa M\'arka and Szabolcs M\'arka for fruitful discussions. C.C.Y. and P.M. acknowledge support from the Eberly Foundation. 
The work of K.M. is supported by the NSF Grant No.~AST-1908689, No.~AST-2108466 and No.~AST-2108467, and KAKENHI No.~20H01901 and No.~20H05852. 
A.P. is supported by the European Research Council via ERC consolidating grant 773062 (acronym O.M.J.). I.B. acknowledges the support of NSF under awards PHY-1911796 and PHY-2110060 and the Alfred P. Sloan Foundation.

\bibliographystyle{aasjournal}
\bibliography{ref}

\end{document}